%% file: main.tex
\documentclass[conference]{IEEEtran}
\IEEEoverridecommandlockouts

\usepackage{cite}
\usepackage{amsmath,amssymb,amsfonts}
\usepackage{algorithmic}
\usepackage{graphicx}
\usepackage{textcomp}
\usepackage{xcolor}
\usepackage[hyphens]{url}

\usepackage{caption} 

\DeclareCaptionFont{my8pt}{\fontsize{8pt}{9.6pt}\selectfont}

\usepackage[normalem]{ulem}
\usepackage{url}
\usepackage{tikz}
\usepackage{amsthm}
\usepackage{amsmath,amsfonts}
\usepackage{graphicx}
\usepackage{textcomp}
\usepackage[table,xcdraw]{}
\usepackage{booktabs} 
\usepackage{stfloats}
\usepackage{wrapfig}
\usepackage{subfig}
\usepackage{multirow}
\usepackage{array}
\usepackage{paralist}
\usepackage{balance}
\usepackage{bm}
\usepackage{pifont}
\usepackage{subcaption}
\usepackage{tabularx}

\usepackage{xspace}
\usepackage{pifont}
\usepackage{algorithmic}
\usepackage[ruled,linesnumbered]{algorithm2e}
\usepackage{diagbox}
\usepackage{threeparttable}
\usepackage{makecell}
\usepackage{colortbl}
\usepackage{bm}
\definecolor{grey}{RGB}{0.5,0.5,0.5}
\usepackage{tcolorbox}
\usepackage[dvipsnames]{xcolor}
\usepackage{lipsum}
\usepackage{enumitem}

\def\BibTeX{{\rm B\kern-.05em{\sc i\kern-.025em b}\kern-.08em
    T\kern-.1667em\lower.7ex\hbox{E}\kern-.125emX}}
\begin{document}

\pdfpagewidth=8.5in
\pdfpageheight=11in

\newcommand{\iscasubmissionnumber}{NaN}

\pagenumbering{arabic}

\title{KVNAND: Efficient On-Device Large Language Model Inference Using DRAM-Free In-Flash Computing}

\author{
    \IEEEauthorblockN{
        Lishuo Deng, Shaojie Xu, Jinwu Chen, Changwei Yan, \\
        Jiajie Wang, Zhe Jiang, and Weiwei Shan
    }
    \IEEEauthorblockA{
        Southeast University \\
        Nanjing, China \\
        Email: \{dengls, 220246825, 230228386, yancw, 220256672, 101013615, wwshan\}@seu.edu.cn
    }
}


\maketitle
\thispagestyle{plain}
\pagestyle{plain}


\begin{abstract}

Deploying large language models (LLMs) on edge devices enables personalized agents with strong privacy and low cost. However, with tens to hundreds of billions of parameters, single-batch autoregressive inference suffers from extremely low arithmetic intensity, creating severe weight-loading and bandwidth pressures on resource-constrained platforms. Recent in-flash computing (IFC) solutions alleviate this bottleneck by co-locating weight-related linear computations in the decode phase with flash, yet still rely on DRAM for the key–value (KV) cache. As context length grows, the KV cache can exceed model weights in size, imposing prohibitive DRAM cost and capacity requirements. Attempts to offload KV cache to flash suffer from severe performance penalties.

We propose KVNAND, the first DRAM-free, IFC-based architecture that stores both model weights and KV cache entirely in compute-enabled 3D NAND flash. KVNAND addresses the fundamental performance challenges of flash under intensive KV cache access by leveraging IFC for all memory-bound operations to reduce data transfer overhead, introducing head-group parallelism to boost throughput, and employing page-level KV cache mapping to align token access patterns with flash organization. In addition, we propose a design space exploration framework that evaluates discrete and compact KVNAND variants to balance weight and KV placement, automatically identifying the optimal design trade-off. These techniques mitigate latency, energy, and reliability concerns, turning flash into a practical medium for long-context KV storage. Evaluations on MHA 7B and GQA 70B LLMs show that KVNAND achieves 1.98×/1.94×/2.05× geomean speedup at 128/1K/10K-token contexts compared to DRAM-equipped IFC designs and addresses out-of-memory failures at 100K context length.


\end{abstract}

\input{contents/1-intro}

\input{contents/2-background}

\input{contents/3-motivation}

\input{contents/4-KVNAND_design}

\input{contents/5-evaluation}

\input{contents/6-related_work}
\input{contents/7-conclusion}

\bibliographystyle{IEEEtranS}
\bibliography{ISCA2026}

\end{document}

%% file: contents/1-intro.tex
\section{Introduction}
\label{sc:Intro}

As Large Language Models (LLMs) integrate into daily workflows, demand increases for personalized AI agents that align with user preferences, domain knowledge, and interaction styles. Deploying such agents on edge devices offers privacy, low-latency responsiveness, and cost efficiency by eliminating cloud dependency, making on-device LLMs a compelling direction for AI democratization  \cite{yu_cambricon-llm_2024}.


Realizing high-quality personal LLM agents on resource-limited edge devices faces two main bottlenecks: memory capacity and bandwidth. \cite{sun_lincoln_2025,yu_cambricon-llm_2024,zhou_survey_2024,xu_-device_2024} Modern LLMs require tens to hundreds of GBs just for weights (e.g., LLaMA2-70B needs \(\sim 140\) GB \cite{touvron_llama_2023-1}), far exceeding typical edge-device DRAM limits (8–16 GB on smartphones, $<$64 GB on laptops). The growing demand for long-context agentic workflows like long document analysis \cite{koh2022empirical}, multi-turn dialogue \cite{zhang2025survey}, and chain-of-thought reasoning \cite{chen2025towards} introduces the KV cache as another dominant consumer of this limited memory \cite{fu2025h2eal,wang2025dynakv}. Moreover, recent state-of-the-art (SoTA) models support extensive context lengths ranging from 128K (LLaMA3.1-70B \cite{llamateam_llama_2024}) to 1M (Gemini 2.5 Pro \cite{comanici2025gemini}). The KV cache demand scales linearly with context length; for example, a 13B model already requires \(\sim 8\) GB KV memory at a 10K context \cite{touvron_llama_2023-1}, placing prohibitive pressure on edge resources. Meanwhile, unlike the compute-intensive prefill stage dominated by matrix–matrix multiplication (GEMM), decoding reduces to memory-bound matrix–vector multiplication (GEMV), exhibiting extremely low arithmetic intensity (\(\approx1 \) OPS/Byte at FP16).

\begin{figure}[tb]
    \centering
    \includegraphics[width=1.0\linewidth, trim=20 30 20 30, clip]{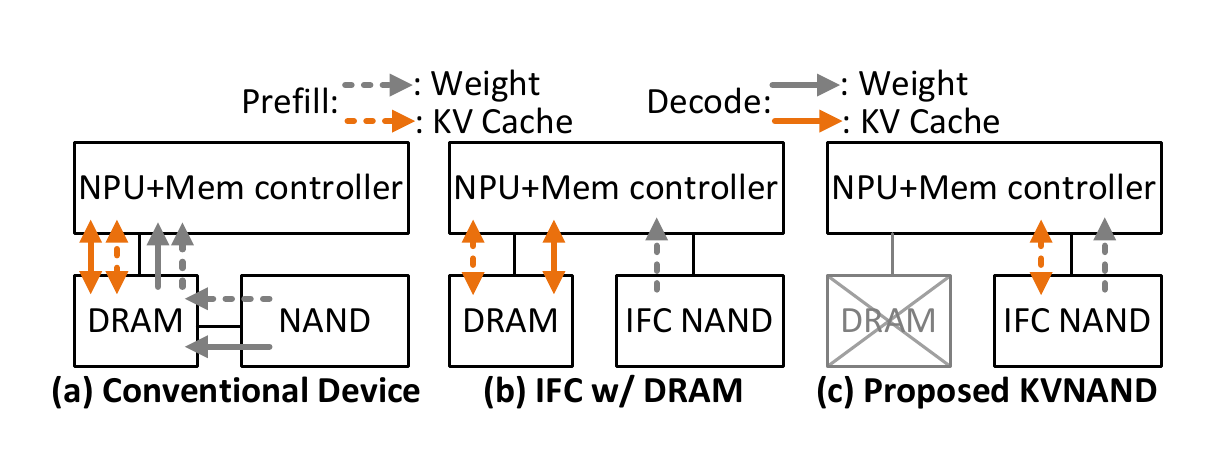}
    \caption{Different on-device LLM inference architectures.}
    \label{fig:intro}
\vspace{-20pt}
\end{figure}

To address the challenges of memory footprint, conventional devices selectively load weights from high-capacity flash into DRAM\cite{alizadeh_llm_2024}, but still suffering from limited flash I/O bandwidth. Figure ~\ref{fig:intro}(b) summarizes more advanced in-flash computing (IFC) designs such as Cambricon-LLM \cite{yu_cambricon-llm_2024} and Lincoln \cite{sun_lincoln_2025}, which push memory-bound operations into the flash to utilize the higher inner bandwidth. Unfortunately, these designs still rely on DRAM for the KV cache, which imposes prohibitive cost and energy under long contexts, while KV-dependent attention computations remain on the NPU, constraining performance.

To overcome these bottlenecks, we propose KVNAND, a DRAM-free NPU-IFC architecture that unifies model weights and KV cache within compute-enabled 3D hybrid-bonding NAND flash. KVNAND eliminates the need for DRAM by storing KV cache directly in flash, leveraging the large capacity and low cost. The architecture extends IFC coverage beyond model weights-related GEMVs to the KV-related GEMVs to improve performance. This approach ensures that all memory-bound computations are serviced by IFC, avoiding intensive KV cache transfers.

To accommodate diverse model scales and context length requirements, we design two complementary KVNAND variants and explore the design space, drawing inspiration from the fundamental tradeoffs inherent in pipeline-parallelism (PP) and tensor-parallelism (TP). KVNAND-Discrete places weights and KV cache in separate groups of IFC dies for KV-heavy cases, enabling head-group pipeline parallel execution that overlaps QKV generation with attention computation. KVNAND-Compact instead stores KV cache in-place within the same flash arrays as the weights for short context length, exploiting higher TP under the same hardware budget and reducing KV inter-die communication. 

KVNAND further introduces a page-level KV cache mapping strategy that aligns KV placement with flash’s page-level access granularity and the distinct access patterns of KV generation and attention. With the aid of lightweight KV buffers, this scheme improves read/write efficiency and significantly reduces redundant page reads. In addition to IFC's core benefit of significantly reducing the movement of intermediate results and large-batch KV cache, the external bandwidth of modern NAND is rapidly evolving. Per-die bandwidth has increased from 3.6 GB/s (ONFI 5.2) to 4.8 GB/s \cite{OnfiSpecs, m31_technology_corporation_onfi_nodate,yanagidaira20251tb}, closely approaching that of LPDDR5X DRAM (8 GB/s) \cite{micron_technology_inc_automotive_2025}  Reliability concerns can also be alleviated by using periodic reclaim \cite{sun_lincoln_2025} and intensive spare flash capacity.

To the best of our knowledge, KVNAND is the first on-device LLM inference architecture that offloads all storage-intensive operands directly into compute-enabled flash dies. Our design preserves the key advantages of SoTA IFC-based LLM accelerators while further extending the role of modern 3D NAND flash as both the storage substrate and execution engine for deploying long-context LLMs on edge devices. The main contributions are as follows:

\noindent
\begin{itemize}[leftmargin=18pt,itemsep=3pt,topsep=3pt]
\item  We propose the first DRAM-free IFC-NPU heterogeneous on-device LLM inference accelerator, which 
stores both the KV cache and weights and executes memory-bound operations in compute-enabled 3D NAND flash.

\item We introduce discrete and compact variants of KVNAND and provide an design space exploration (DSE) framework for selecting configurations under given model, context length, and hardware constraints.

\item We optimize attention layer dataflow and KV cache mapping by aligning 
LLM inference characteristics with KVNAND flash storage properties to reduce execution time and energy while alleviating reliability stress.
\item 
We evaluate the performance and power of KVNAND. Across MHA 7B and GQA 70B LLMs, KVNAND achieves 1.98\(\times\)/1.94\(\times\)/2.05\(\times\) geomean speedup compared with DRAM-equipped IFC designs under 128/1K/10K-token contexts, and 1.17\(\times\)/1.32\(\times\) geomean energy efficiency improvements under 10K/30K-token contexts. In addition, KVNAND reduces memory cost by 69\% and solves the out-of-memory at 100K context length.
\end{itemize}

%% file: contents/2-background.tex
\section{Background}
\label{sc:Background}

\subsection{LLM Inference}
\begin{figure}[tb]
    \centering
    \includegraphics[width=1.0\linewidth, trim=20 25 20 25, clip]{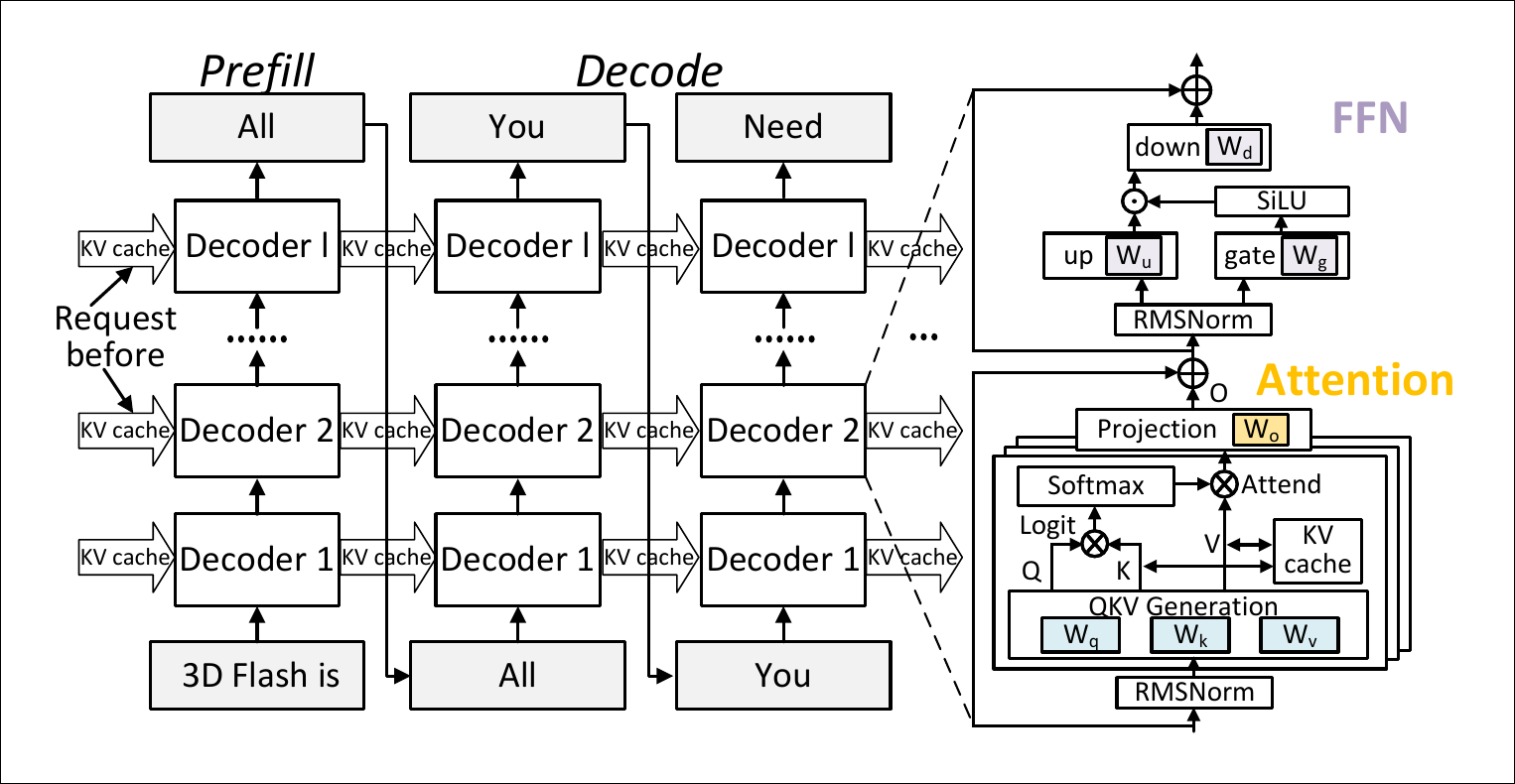}
    \caption{General LLM architecture and inference flow.}
    \label{fig:background}
\vspace{-10pt}
\end{figure}

Mainstream decoder-only LLMs \cite{zhang_opt_2022,touvron_llama_2023,jiang_mixtral_2024} consist of stacked decoder blocks, each with an attention module and a feed-forward network (FFN) as shown in Figure ~\ref{fig:background}. The attention layer generates Q/K/V, performs self-attention (Logit, Softmax, Attend), while fully connected (FC) layers dominate the FFN. Multi-head attention (MHA) \cite{vaswani_attention_2023} enables parallel attention across different aspects of the context. Group-query attention (GQA) \cite{ainslie_gqa_2023} is an efficient attention architecture that strikes a balance between the high quality of MHA and the low inference cost of Multi-Query Attention (MQA) \cite{shazeer2019fast} by grouping multiple query heads to share a single K and V head.

For a single inference request, the input sequence is first processed in prefill phase to generate an initial output token, followed by decode phase, where tokens are generated one at a time. A widely adopted technique is KV cache \cite{kwon_efficient_2023}, lowering attention complexity from \(O(s^2)\) to \(O(s)\). However, as context length grows, KV cache becomes both capacity-intensive and bandwidth-demanding, especially since conversational history and reasoning traces further extend effective context \cite{lee_infinigen_2024,openai_introducing_2025, deepseek-ai_deepseek-r1_2025}. To relieve DRAM pressure, recent GPU-based inference architecture offloads KV caches to flash \cite{sheng_flexgen_2023,rasley_deepspeed_2020,pan_instattention_2025}, enabling larger contexts on memory-constrained devices. Yet, even with SSD-side compute\cite{pan_instattention_2025}, these designs fail to exploit flash’s internal bandwidth and remain impractical for resource-constrained consumer devices.

As shown in the roofline model of Figure ~\ref{fig:motivation_2}(a), decode-phase GEMV computations are memory-bound, with performance constrained by the limited bandwidth of mobile NPUs. In single-batch on-device LLM inference, the absence of batch-level parallelism eliminates weight reuse across tokens, further lowering arithmetic intensity \cite{he_papi_2025,yu_cambricon-llm_2024,pan_instattention_2025}. Consequently, data movement accounts for over 90\% of decode latency \cite{sun_lincoln_2025}, leaving compute units heavily underutilized.


\subsection{In-Flash Computing for LLM}
To address the low arithmetic intensity of LLM decoding, recently, Processing-In-Memory (PIM) has emerged as a promising direction \cite{heo_neupims_2024,gu_pim_2025,he_papi_2025,li_h2_2025}. In particular, on-die in-flash computing (IFC) leverages the low cost and high storage density of 3D NAND flash to perform near-data processing \cite{li_large_2025,chun_pif_2022}. Recent works demonstrate IFC effectiveness for on-device LLMs, achieving up to 36.3 tokens/s on LLaMA2-7B \cite{yu_cambricon-llm_2024} and 11.5 tokens/s on LLaMA-65B \cite{sun_lincoln_2025}, sufficient for real-time inference. Moreover, the high internal bandwidth and large capacity offered by 3D stacking technologies further accentuate the advantages of IFC for on-device LLM accelerators.

Cambricon-LLM \cite{yu_cambricon-llm_2024} adopts a chiplet-based hybrid architecture with compute-enabled flash dies connected to the NPU via high-speed die-to-die links. Lincoln \cite{sun_lincoln_2025} integrates compute logic beneath each flash plane and interfaces flash to the SoC through LPDDR PHYs. AiF \cite{lee_aif_2025} further optimizes flash circuits and reliability. These advances define a new architectural paradigm for on-device LLM inference, referred to as IFC-NPU. IFC subsystems accelerate memory-bound operations involving model weights in the decode phase. The NPU handles the prefill phase and nonlinear computations. In these architectures, flash is used solely for storing weights, with KV caches maintained in DRAM.

%% file: contents/3-motivation.tex
\section{Motivation}
\label{sc:Motivation}




\begin{figure}[tb]
    \centering
    \includegraphics[width=1.0\linewidth]{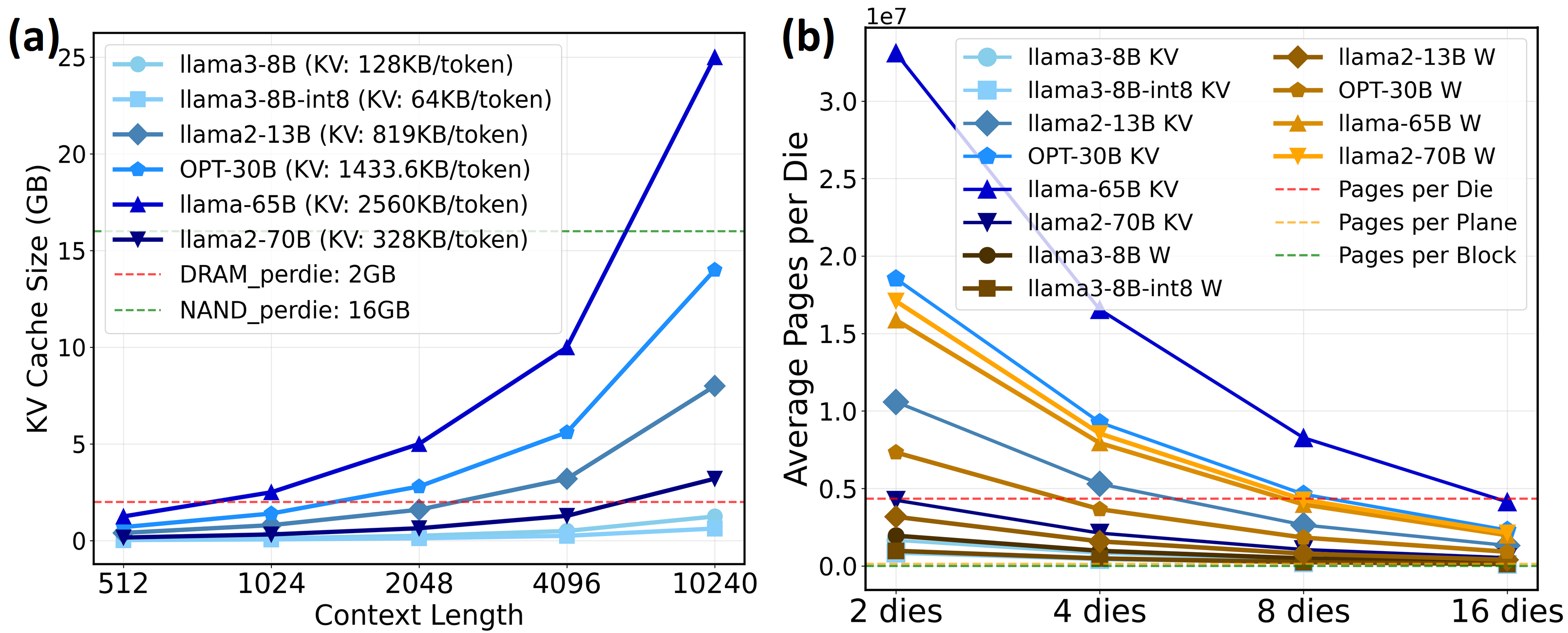}
    \caption{(a) KV cache size of various models as the context length increases. (b) Number of NAND flash pages occupied by KV cache and model weights.}
    \label{fig:motivation}
\vspace{-5pt}
\end{figure}

\begin{figure}[tb]
    \centering
    \includegraphics[width=1.0\linewidth, trim=30 20 30 20, clip]{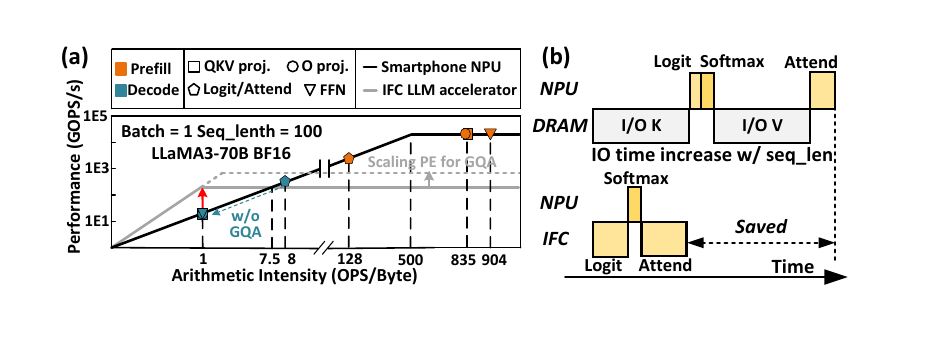}
    \caption{(a) Roofline model analysis of IFC and smartphone NPU. (b) Comparison of attention computation workflows when KV cache is stored in DRAM and in NAND with IFC.}
    \label{fig:motivation_2}
\vspace{-10pt}
\end{figure}

\subsection{The Overlooked Crisis: Long Context KV Cache Storage} 
With the rapid growth of context lengths in LLM inference, the KV cache scales with sequence length and can surpass model weights as the dominant consumer of memory \cite{jin_llm_2024,lin_infinite-llm_2024}. The problem is especially severe for large-scale models preserving long conversational histories. Figure ~\ref{fig:motivation}(a) illustrates that KV cache size grows rapidly with context length and can even exceed model weights. For instance, with a 10K context, LLaMA2-70B requires several gigabytes of KV storage despite using GQA. Figure ~\ref{fig:motivation}(b) further shows the flash page occupancy of 16GB flash dies \cite{sun_lincoln_2025}, where large models such as LLaMA2-70B demand multiple dies exclusively for KV cache at 100K tokens, while smaller GQA-based models like LLaMA3-8B \cite{llamateam_llama_2024} also needs one to two 16GB IFC dies. Compared to DRAM, 3D NAND flash provides the same capacity at about one-fifth the cost \cite{yu_cambricon-llm_2024} and is more resilient to out-of-memory (OOM) failures. These trends highlight flash as a practical and scalable medium for long-context KV storage in on-device LLM inference.


\begin{figure}[tb]
    \centering
    \includegraphics[width=1.0\linewidth]{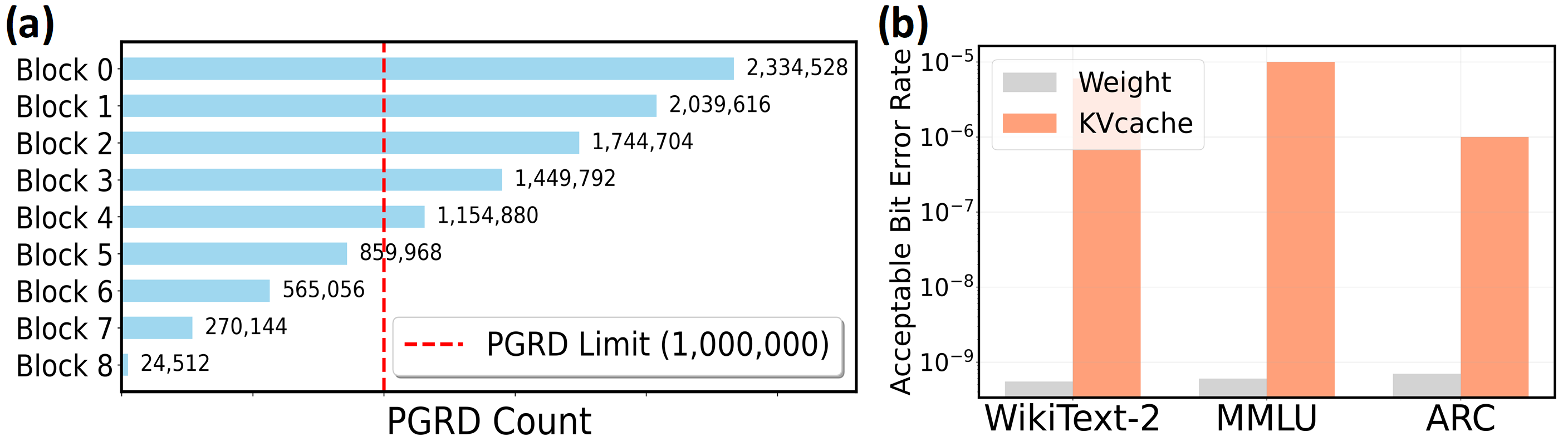}
    \caption{(a) Page-read (PGRD) counts for LLaMA3.1-8B KV cache with a 50K-token context (1K input and 1K generated tokens per request). (b) Acceptable BER in model weights and KV cache resulting in a 10\% accuracy/perplexity degradation. }
    \label{fig:motivation_3}
\vspace{-10pt}
\end{figure}

\subsection{The Latency Barrier: Why Naive NAND Substitution Fails}
A naive substitution of NAND for DRAM to offload the KV cache, while addressing the aforementioned capacity and cost issues, introduces significant performance challenges due to the higher latency of NAND flash.
First, flash's external bandwidth trails that of DRAM. For per-die external bandwidth, a single LPDDR5X channel can sustain about 8 GB/s \cite{micron_technology_inc_automotive_2025}, whereas modern NAND flash (even ONFI 6.0) reaches 4.8 GB/s \cite{m31_technology_corporation_onfi_nodate}.
More critically, the internal access latency gap is fundamental. DRAM provides access at the nanosecond (ns) scale. Conversely, NAND's internal program operations are orders of magnitude slower, operating at the microsecond (µs) scale (e.g., a conservative 75 µs program time \cite{kouchi_135_2020}, 1.6 GB/s internally).
This stark latency divide renders simple offloading infeasible. It necessitates specialized KV cache storage and mapping schemes that align with flash access characteristics, thereby hiding fragmented access patterns and mitigating their latency impact. This aspect will be elaborated in Section \ref{sc:KVNAND Design}.


Here, we first conduct an evaluation of a naive design that simply replaces DRAM with flash in Lincoln's architecture \cite{sun_lincoln_2025}, with an internal read bandwidth of 32 GB/s, assuming ideal KV cache access and 4 dies in parallel. This bandwidth level analysis provides an initial understanding of the feasibility and limitations of storing KV cache in flash. We use the popular mixtural-of-experts (MoE) model Mixtral-8\(\times\)7B \cite{jiang_mixtral_2024} as a case study. In the single-batch decode phase, generating one token with GQA yields a KV cache size of: \({KV\_per\_tk} = 2\times l\times k \times \frac{d}{h} \times 2B = 128KB\) in BF16, where the factor of 2 accounts for both K and V tensors, \(l\) is the number of decoder block layers, \(d\) is the hidden size.



(1) Reading KV cache from flash. For larger KV cache reads, the flash page buffer allows pipelining, which hides part of the access latency. During attention computation with a sequence length of 1K tokens, the read latency is: \(t_{\text{KV,read}} = ({KV\_per\_tk} \times seq\_length) / (4 \times {BW_\text{external}^\text{read}}) \approx 6.9\,ms\).



(2) FFN acceleration in IFC. Each routed expert has about 175M parameters, or 87.5 MB under INT4 quantization. With 8 experts per layer and 32 layers, the FFN weights total roughly 22.4 GB, but only 2 experts are activated per inference step. The corresponding read time is: 
\(t_{\mathrm{FFN,read}} = (l \times {expert\_size} \times {num\_expert}) / (4 \times {BW_\text{internal}^\text{read}}) \approx 44\,ms\).



For short contexts, FFN execution dominates end-to-end latency, with attention contributing only a minor fraction. As the context length grows, attention latency increases proportionally and the bandwidth gap between flash and DRAM becomes more pronounced. Similar trends hold across different LLM scales. Consequently, an intuitive solution to this KV movement latency is to increase the external NAND bandwidth. For instance, Lincoln \cite{sun_lincoln_2025} connects IFC dies via LPDDR-style interfaces, largely mitigating the external I/O bottleneck, while other proposals leverage high-speed custom die-to-die links \cite{yu_cambricon-llm_2024}. However, these specialized solutions incur significant custom design costs and forego the benefits of leveraging the mature and commercial ONFI interface.

\subsection{Unleashing Full Potential: IFC for Mem-Bound Operands}
As our analysis confirms, transferring large volumes of KV cache from DRAM to the NPU incurs significant latency as sequence length increases. Rather than merely optimizing this costly I/O path, a more fundamental solution is to avoid the data movement altogether. This motivates our proposal to leverage an IFC architecture, as shown in Figure ~\ref{fig:motivation_2}(b). All three SoTA IFC-based LLM inference accelerators \cite{yu_cambricon-llm_2024,sun_lincoln_2025,lee_aif_2025} leverage the insight from the roofline analysis in Figure ~\ref{fig:motivation_2}(a): the memory-bound operands of LLM inference should be offloaded to IFC if possible. However, in these designs, KV cache remains stored in DRAM, which forces the memory-bound Logit and Attend operands in attention layer to be executed on NPU. In contrast, storing KV cache directly in NAND allows these two operations to be executed entirely within IFC die, saving associated I/O time and energy. This enables all memory-bound operators in LLM inference to be offloaded to IFC units. 


Noted that models with a GQA architecture exhibit an \(h/k\times\) (typically 4 or 8) arithmetic intensity, pushing it beyond the roofline intersection and making it less memory-bound, because the number of KV heads \(k\) is smaller than the number of Q heads \(h\). Therefore, for an IFC system to maintain a performance benefit on GQA workloads, its peak computational throughput must be correspondingly increased as illustrated by the gray dashed-line in Figure ~\ref{fig:motivation_2}(a). As shown in Section~\ref{sc:eval}, the resulting workload remains well-suited for IFC execution in most cases.

\subsection{Reliability Consideration}
The reliability problems like retention, read disturbance, and endurance of NAND flash \cite{zhao_error_2023,cai_error_2013}, have been explicitly considered in several flash-based LLM accelerators \cite{yu_cambricon-llm_2024,sun_lincoln_2025,lee_aif_2025,pan_instattention_2025}. Lincoln identified read disturbance \cite{cai_read_2015,zhang_novel_2017,kim_reliability_2020} as the primary reliability concern that increases the bit error rate (BER) since the continuous long-output generation. For weight storage, each page is written once and then read many times for decoding, leaving sufficient programs/erases (P/E) endurance margin to perform read reclaim operations \cite{kim_behemoth_2021}. 




For KV cache stored in NAND flash, two primary reliability stress factors must be considered: (1) increased P/E cycles due to KV updates, and (2) accumulated page reads. The access pattern of the KV cache differs from that of model weights, as KV entries generated during decoding are only consumed in subsequent token-generation iterations. As shown in Figure \ref{fig:motivation_3}(a), under the flash configuration of Lincoln \cite{sun_lincoln_2025} with a 50K-token context window (1K input and 1K generated tokens per request), blocks storing early KV entries exhibit higher cumulative page-read counts, while blocks holding later KV caches remain well below the read-disturb endurance limit. Fortunately, flash capacity enables trading space for reliability by retiring high-BER blocks. Moreover, fragmented KV accesses can amplify page reads, a challenge we mitigate through the mapping schemes in Section \ref{sc:KVNAND Design}.

Bit-error injection further validates the feasibility. As illustrated in Figure \ref{fig:motivation_3}(b), experiments on LLaMA3.1-8B \cite{llamateam_llama_2024} across WikiText-2\cite{merity_pointer_2016}, 
MMLU\cite{ hendrycks_measuring_2021}, 
ARC\cite{clark_think_2018} demonstrate that accuracy degrades by over 10\% at much lower BER thresholds for weights than for KV cache, indicating that KV data are inherently more error-tolerant due to the error-masking effect of the Softmax operation \cite{xie_realm_2025}.


%% file: contents/4-KVNAND_design.tex
\section{KVNAND Design}
\label{sc:KVNAND Design}




\begin{figure}[tb]
    \centering
    \includegraphics[width=1.0\linewidth, trim=18 18 44 18, clip]{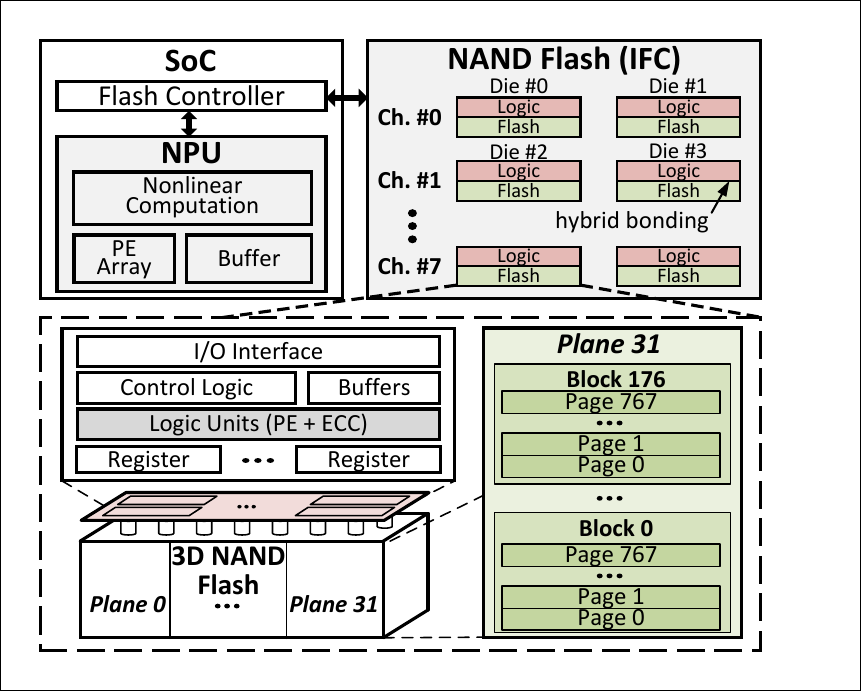}
    \caption{KVNAND overall architecture.}
    \label{fig:KVNAND_design}
    \vspace{-10pt}
\end{figure}


\subsection{Overall Architecture and Dataflow}

\textbf{Hardware design.} 
Figure \ref{fig:KVNAND_design} illustrates the core concept of the proposed KVNAND architecture: both model weights and KV cache are stored and computed inside compute-enabled flash, enabling fully DRAM-free inference. By leveraging IFC beyond weight-only GEMV to KV-related operands, KVNAND eliminates intensive KV transfers in a long context window. The system consists of 3D NAND flash dies with IFC, a neural processing unit (NPU), and a multi-channel flash controller. The NPU and flash controller are integrated into the system-on-chip (SoC).

\begin{figure}[tb]
    \centering
    \includegraphics[width=1.0\linewidth, trim=0 10 0 10 clip]{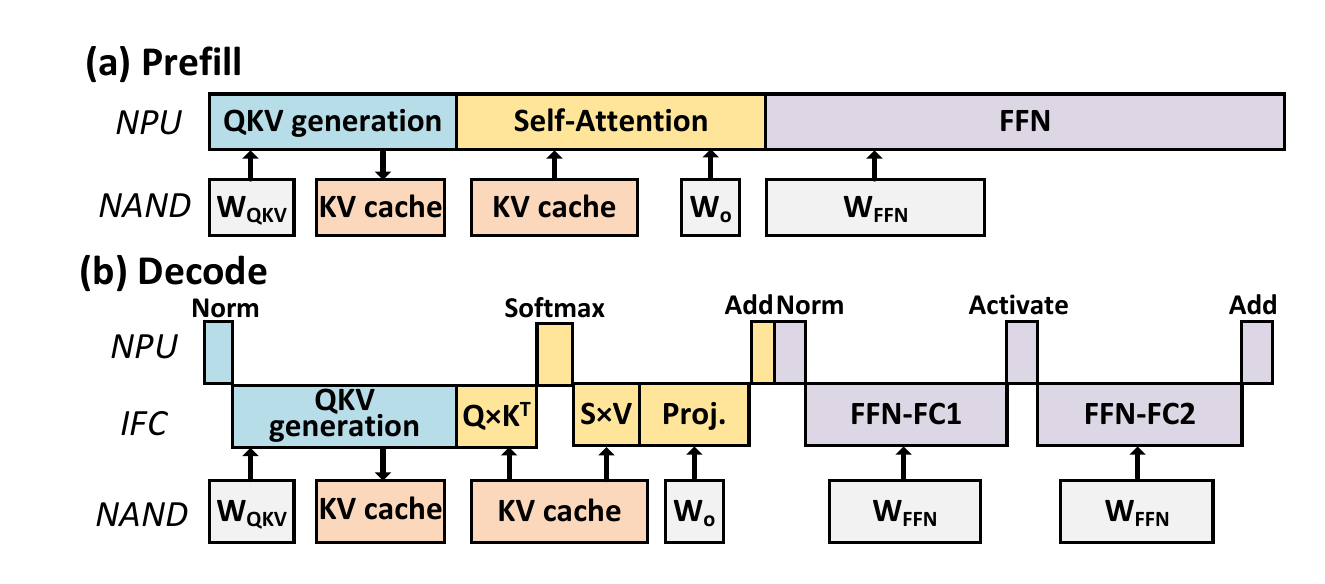}
    \caption{Overall dataflow of (a) Prefill and (b) Decode phase.}
    \label{fig:KVNAND_design_2}
    \vspace{-10pt}
\end{figure}

In KVNAND architecture, 3D NAND flash arrays and their peripheral circuits are connected through wafer-to-wafer hybrid bonding \cite{sun_lincoln_2025,moschiano_tutorial_2024,ouyang_excellent_2021,yolanda_wafer_2022}. On the upper CMOS die originally reserved for flash peripheral circuitry, additional logic units with processing elements (PEs) are integrated to form the logic die
, enabling GEMV operations between stored weight matrices or KV cache and input vectors. Data read from the flash array first enter the data and cache registers, passes through lightweight on-die ECC, and is then fed into the PE. Computation results and input vectors are exchanged between IFC and NPU as needed.

The NPU is implemented on a separate die and integrates general-purpose PE arrays for high-throughput GEMM, residual connection (add), as well as specialized functional units for operations such as normalization, rotary position embedding (RoPE), activation, and Softmax. These vector operations are ill-suited for IFC, as they offer no bandwidth savings due to identical input/output sizes and incur prohibitive design complexity within the resource-constrained logic die. Moreover, the distributed outputs from multiple parallel PEs in IFC must be aggregated before certain computations. 

The flash controller on the SoC manages communication with multiple IFC dies, each connected through an independent channel. Multiple channels enable scaling of aggregate I/O bandwidth and computation parallelism across IFC dies.




\begin{figure*}[tb]
    \centering
    \includegraphics[width=1.0\linewidth, trim=25 20 45 20, clip]{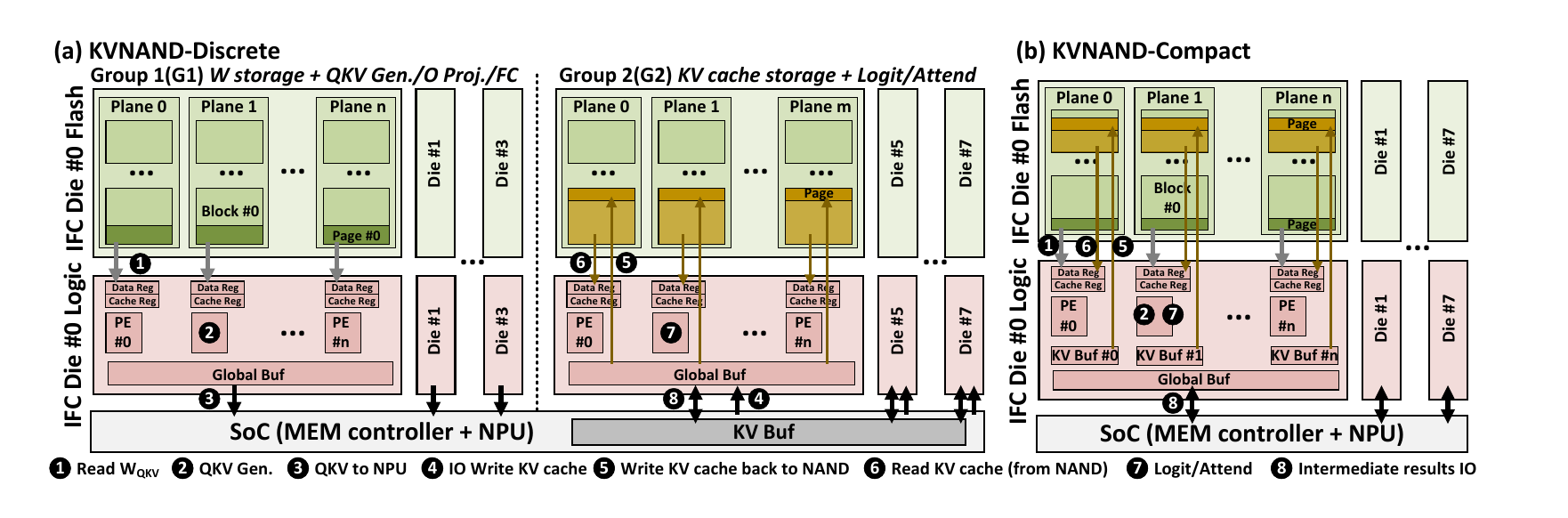}
    \caption{(a) KV-Discrete and (b) KV-Compact architectures with the QKV-Gen. and Attention executing flow in decode phase.}
    \label{fig:KVNAND_design_3}
    \vspace{-10pt}
\end{figure*}

\textbf{Compute Flow.} Figure \ref{fig:KVNAND_design_2}(a) depicts the prefill phase, which is dominated by compute-bound GEMM. Prefill is therefore executed on the NPU, following prior schedulers \cite{yu_cambricon-llm_2024,sun_lincoln_2025,lee_aif_2025}. The NPU streams weights from flash, performs QKV projection, self-attention, and the FFN. The resulting KV tensors are stored in flash.

Figure \ref{fig:KVNAND_design_2}(b) illustrates the decode phase, where KVNAND fully exploits the advantage of storing the KV cache in flash. All memory-bound GEMVs of LLM single-batch inference are offloaded to the IFC logic die. This includes QKV generation, Logit, Attend, O projection, and the FC layers in the FFN. In this phase, the NPU is only responsible for nonlinear functions. Both phases require bidirectional data exchange between the NPU and flash, with intermediate data and KV cache updates passing through the interface. Unlike prior IFC accelerators that use DRAM to buffer embeddings and intermediate results, KVNAND keeps these smaller data structures in on-chip SRAM.
Some works \cite{lee_aif_2025,yu_cambricon-llm_2024} offload part of the decode-phase GEMV computation to NPU to balance utilization, but such techniques require additional system-level support. KVNAND establishes a baseline architectural optimization that can be seamlessly combined with these methods.

\subsection{KVNAND-Discrete and KVNAND-Compact}\label{sec:KVNAND-Discrete and KVNAND-Compact}
When both weights and KV cache are stored in flash and attention GEMV operations are executed on IFC, the design space becomes more complex than in DRAM-equipped designs \cite{yu_cambricon-llm_2024,sun_lincoln_2025,lee_aif_2025}. We propose two design variants, KVNAND-Discrete (KVNAND-D) and KVNAND-Compact (KVNAND-C), which are shown in Figure \ref{fig:KVNAND_design_3}. These two variants are explicitly designed to accommodate diverse model scales or context lengths.

As shown in Figure \ref{fig:KVNAND_design_3}(a), KVNAND-D partitions all IFC dies into two groups: Group 1 (G1) stores model weights and executes QKV generation,  O projection and FC, reusing dataflow of prior IFC  \cite{sun_lincoln_2025}, while Group 2 (G2) stores KV cache. Both groups share the same die configuration for cost-effective manufacturing, while the die allocation between G1 and G2 can be flexibly tuned. For short contexts, FFN dominates latency, favoring more dies in G1. As context length grows, attention latency and KV cache size increase, making additional G2 dies more beneficial. The configuration trade-offs will be further discussed in Section \ref{sc:eval}.

The execution dataflow involving KV cache in KVNAND-D is as follows. \(W_{QKV}\) is partitioned in G1 to increase per-head generation parallelism as in Lincoln \cite{sun_lincoln_2025}. Each page of multi-plane is read simultaneously (\ding{182}) and multiplied with an input vector broadcast to all planes across all dies in the PE (\ding{183}), with partial results accumulated in the PE registers. When a complete output is produced, it is collected in the global buffer and sent to the SoC as Q/K/V results (\ding{184}). Newly generated KV cache entries are first stored in SRAM-based KV buffer in SoC, and once the accumulated size reaches a proper size,
they are written to G2 flash (\ding{185}\ding{186}). For attention, G2 IFC dies load the required K vectors from flash (\ding{187}) and multiply them with broadcasted Q vectors (\ding{188}). The results are aggregated in the NPU (\ding{189}) to compute the Softmax, which are then sent back to IFC dies to perform the Attend with V vectors (\ding{187}\ding{188}). 

As shown in Figure \ref{fig:KVNAND_design_3}(b), KVNAND-C integrates KV cache and weight storage within the same die, leveraging spare capacity after weight placement to accommodate KV data. In this design, Q/K/V generation also begins by reading the \(W_{QKV}\) pages (\ding{182}) and generating Q/K/V in the PEs (\ding{183}). Generated K and V are temporarily stored in the KV buffer of each plane, and once sufficient KV data is accumulated, it is written back (\ding{186}) to KV-dedicated blocks in the same plane. Self-attention in KVNAND-C begins by loading KV cache from flash (\ding{187}) and, as in KVNAND-D, performing the Logit and Attend in the PEs (\ding{188}), with intermediate results exchanged with the NPU (\ding{189}). 

\begin{figure}[tb]
    \centering
    \includegraphics[width=1.0\linewidth, trim=50 20 20 20, clip]{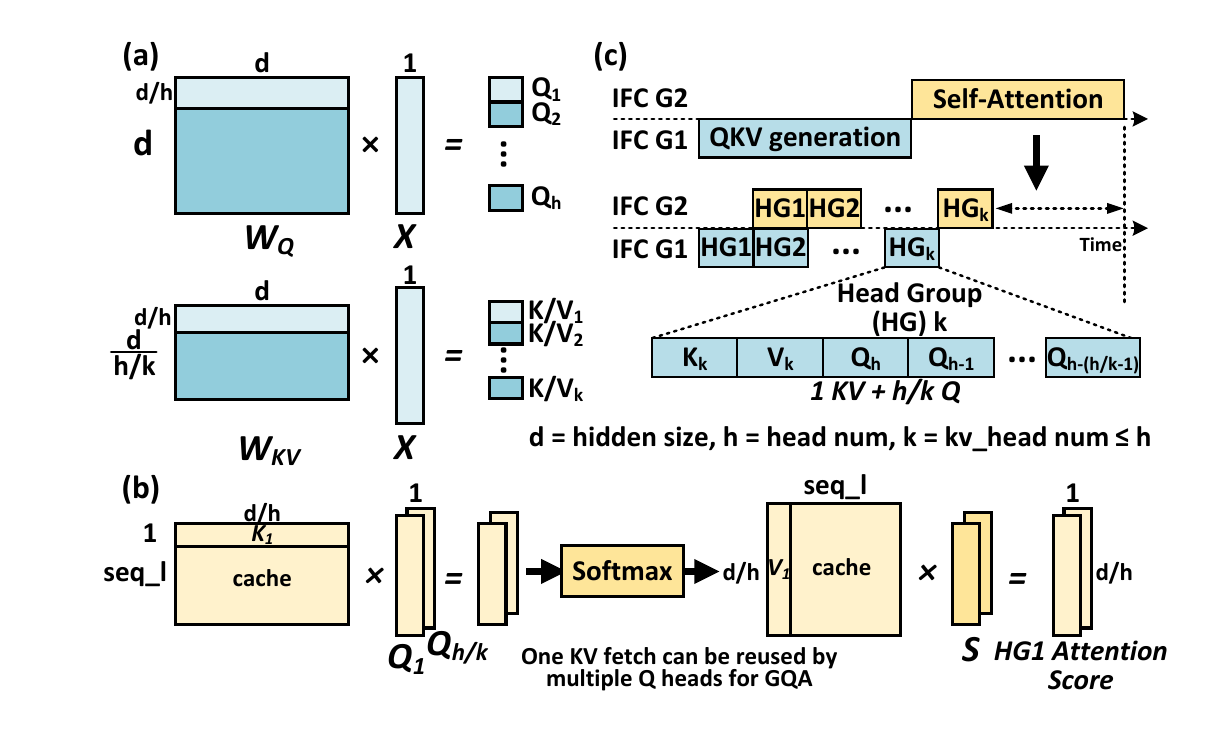}
    \caption{(a) QKV generation and (b) Self-attention of GQA model, and (c) KVNAND-D head-group parallelism dataflow.}
    \label{fig:KVNAND_design_4}
\vspace{-10pt}
\end{figure}

KVNAND-C increases tensor-parallelism by leveraging all hardware resources, whereas KVNAND-D employs pipeline-parallelism to overlap QKV-generation and Logit/Attend but sacrifices some tensor-parallelism. Compared to the discrete design, KVNAND-C also avoids cross-die data movement for KV storage. For an equivalent channel and die configuration, KVNAND-C can offer higher compute parallelism, accelerating QKV generation, attention computation, and the subsequent FFN stage, which often accounts for a large portion of total inference latency when context length is short. Another distinction of the compact design is that ECC encoding must be performed on the logic die to protect the KV data before writing to flash, rather than in the SoC flash controller as in the discrete design. This adds some area and power overhead on the logic die, though the cost is modest since ECC encoding typically requires only about one-quarter of the area of decoding \cite{gao_study_2003,alam_comet_2022,micheloni_inside_2010}.

\subsection{Dataflow Optimization for Attention}
Prior IFC-based accelerators optimized only weight-related GEMVs. With KVNAND storing KV cache in flash, the key is to accelerate attention with KV accesses, as fully utilizing IFC throughput is critical for long-context latency.

Single-batch on-device inference requires strictly sequential decoding for each token, where QKV generation, self-attention, and FFN execute in order. This dependency chain limits intra-layer parallelism. However, QKV computations across heads are independent and can run in parallel, as illustrated in Figure \ref{fig:KVNAND_design_4}. This enables temporal overlap between QKV generation and self-attention, with maximum benefit when attention time is comparable to QKV generation time at a given sequence length. This software-level attention-head partitioning of the LLM decode process naturally aligns with the inherent hardware partitioning of our KVNAND-D architecture. As shown in Figure \ref{fig:KVNAND_design_4}(c), this synergy allows us to leverage the G1 and G2 IFC groups to execute QKV generation and the subsequent attention computation in a decoupled, pipelined fashion. This approach effectively hides the execution latency and minimizes overhead.


 To realize this execution, we introduce the head group (HG) as the fundamental unit of granularity for pipeline-parallelism scheme, as shown in Figure \ref{fig:KVNAND_design_4}(c). Each HG consists of one KV pair and its associated Q heads: for example, in LLaMA3-8B (\(h=32\), \(k=8\)), each HG contains one KV pair and four Q heads, while in LLaMA2-70B (\(h=64\), \(k=8\)), each HG contains one KV pair and eight Q heads. In standard MHA where \(h=k\), each HG naturally reduces to a single QKV triplet. Other parallelism forms, such as concurrent execution of gate and up-projection in FFN, are orthogonal to our approach and can complement in KVNAND.

\begin{figure}[tb]
    \centering
    \includegraphics[width=1.0\linewidth, trim=28 20 20 20, clip]{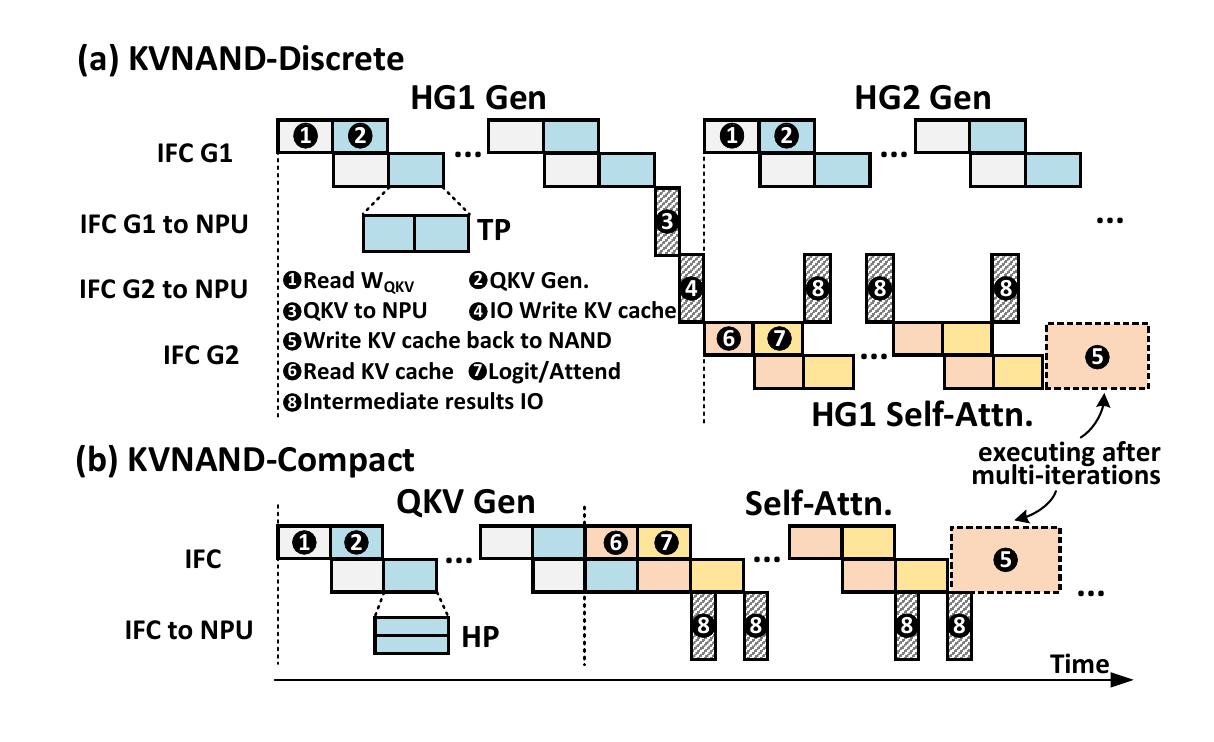}
    \caption{Attention layer sequence flow of (a) KVNAND-D and (b) KVNAND-C.}
    \label{fig:KVNAND_design_5}
\vspace{-10pt}
\end{figure}

Figure \ref{fig:KVNAND_design_5}(a) shows the KVNAND-D timing breakdown with HG-parallelism. HG1 generation leverages the two-stage register pipeline, common in many flash devices, to overlap QKV weight reads (\ding{182}) with GEMV computation (\ding{183}). Following \cite{yu_cambricon-llm_2024,sun_lincoln_2025}, the IFC logic makes \ding{182} and \ding{183} take roughly the same time for bandwidth matching. For instance, if each plane reads a 4 KB page in \(4 \,\mu s\), the compute units are sized to finish the corresponding MACs in \(4 \,\mu s\) (i.e., 2FMACs at 400MHz). Increasing the die count raises parallelism, reducing the number of \ding{182}\ding{183} iterations and thus the total QKV latency. Once computation is completed, results are sent via the IFC die I/O to the NPU memory controller (\ding{184}) and then written into G2 dies (\ding{185}) for self-attention. Because part of the work is already done in IFC, the transfer volume is small and the external bandwidth is negligible. At this point, G1 can immediately start generating HG2, independent of other results. In parallel, G2 performs self-attention GEMV for HG1: KV cache is read (\ding{187}), Logit and Attend computations are executed (\ding{188}) using the same \ding{182}\ding{183} pipeline to overlap part of the read time. For Logit, K is loaded and aggregated on the NPU (\ding{189}), followed by Softmax and broadcasting back to G2 for Attend. The duration of the attention scales with context length, since both KV reads and computation grow with sequence length, as shown in Figure \ref{fig:KVNAND_design_4}(b). A write to the G2 flash (\ding{186}) occurs only after generating multiple tokens, when the KV buffer becomes full, thereby amortizing the write-overhead. 

Note that in GQA models, each KV fetch is reused across multiple Q heads, enabling a single access to serve multiple attention computations. Thus, achieving a bandwidth–compute balance requires scaling the PE count by a factor of $h/k$ to equalize the latencies of phases \ding{182} and \ding{183}. This implies that for a typical GQA model where $h/k=8$, the required bandwidth-compute balance necessitates 16 FMACs per plane operating at 400MHz.
Although this compute capability appears superfluous for weight-related GEMVs executing in G2 or KVNAND-C, it can be exploited for speculative decoding to obtain further performance gains \cite{sun_lincoln_2025}. Moreover, our evaluation in Section \ref{sc:eval} demonstrates that this added compute incurs only a negligible area overhead on the logic die. 

The KVNAND-C timing diagram is shown in Figure \ref{fig:KVNAND_design_5}(b). Unlike KVNAND-D, it cannot overlap HG generation and attention, because both \(W_{QKV}\) reads (\ding{182}) and KV cache reads (\ding{187}) contend for the same flash internal read bandwidth. 
However, this architecture compensates by leveraging the full computational power of all IFC dies in both non-overlapped stages, thereby reducing the iteration count for phases \ding{182}\ding{183} and \ding{187}\ding{188}.
Regarding the detailed dataflow for QKV generation, KVNAND-C requires that QKV head generation be parallelized across HGs (Head-Parallelism, HP). That is, instead of applying all multi-plane parallelism to a single head (Tensor-Parallelism, TP), multiple heads are generated concurrently. This approach resembles superscalar versus scalar execution \cite{kuszmaul_comparison_1999}. While the overall latency remains unchanged given fixed compute resources, this HP execution model is a key prerequisite for the KV cache mapping scheme adopted by KVNAND-C in the next section. In contrast, G1 dies of KVNAND-D can employ TP-based data flow for QKV generation.

\subsection{KV Cache Mapping Scheme}

\begin{figure}[tb]
    \centering
    \includegraphics[width=1.0\linewidth, trim=24 18 28 20, clip]{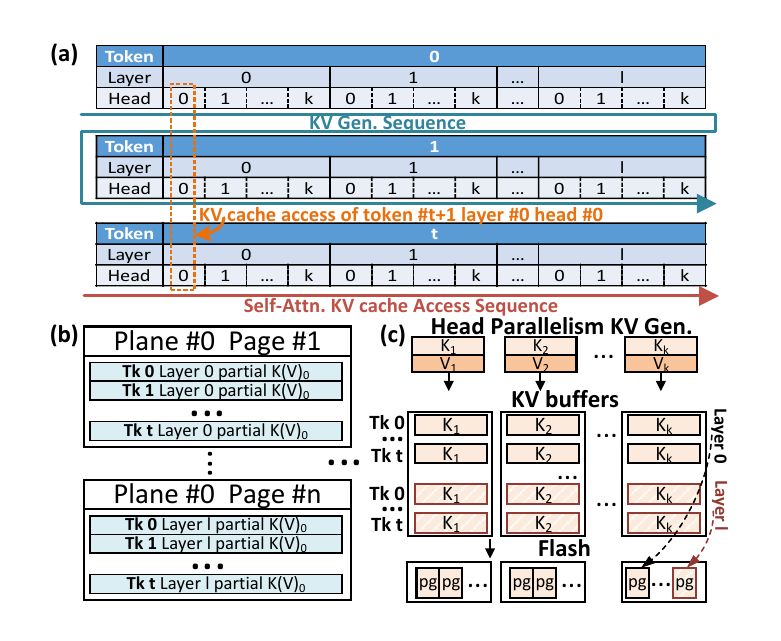}
    \caption{KV cache storage mapping scheme.}
    \label{fig:KVNAND_design_7}
    \vspace{-10pt}
\end{figure}

Unlike model weights, which are read-only during inference and exhibit static access patterns, the KV cache is dynamically generated during decoding and exhibits two unique storage characteristics as described in Figure \ref{fig:background} and Figure \ref{fig:KVNAND_design_7}(a). (1) 
Temporal ordering of token generation, in which KV entries are produced sequentially as new tokens are generated.
(2) Independence of KV usage across layers and heads causes conflicts between fine-grained KV cache operations and coarse-grained flash storage.



\textbf{Page-level KV Cache Mapping.} 
As shown in Figure \ref{fig:KVNAND_design_7}(a), without optimization, KV entries are written in generation order, so each page mixes data from different layers and heads. Since attention requires KV from multiple tokens but only within the same layer and head, this misalignment reduces spatial locality and increases page reads. To avoid this, KV entries belonging to the same layer and head group should be stored contiguously in flash pages, as shown in Figure \ref{fig:KVNAND_design_7}(b). This organization improves page-level locality, increasing the effective hit rate and thus enhancing performance and reliability. A straightforward solution is to buffer KV entries from multiple tokens until one page is filled with data for a single head in a single layer, then write it to flash. However, this requires buffer capacity on the order of \(k \times l \)  pages (roughly \(100\sim 1000\times\) page size), which is infeasible for the limited on-die area in KVNAND-C and would also enlarge the SoC-side buffer in KVNAND-D, potentially reintroducing the need for costly off-chip DRAM.

To address this challenge, we propose a page-level KV cache mapping scheme and its hardware workflow. Taking KVNAND-C as an example,
by parallelizing KV head generation across different planes, as illustrated in Figure \ref{fig:KVNAND_design_7}(c), each plane's KV buffer only needs to hold the KV data of its assigned head until enough tokens are accumulated to fill one page before writing to flash. This reduces the buffer requirement per plane by a factor of \(k\), significantly easing the on-die area constraint. In addition, partial-page updates \cite{kim_subpage_2015,kang_subpage-aware_2018,kawaura_non-volatile_2006} can be applied to write data into a page at a finer granularity, further reducing the required buffer size.

Specifically, under HP generation, the number of tokens (\(t\)) required to fill a page buffer can be expressed as: \(t = page\_size \times (num\_die\times num\_plane / 2k) / KV\_size\_unit\), where \((num\_die\times num\_plane / 2k)\) is the number of planes allocated for generating either K or V after head partitioning, and \(KV\_size\_unit = precision \times \frac{d}{h}\). For FP16 precision, this value is 256B in most mainstream models \cite{zhang_opt_2022,touvron_llama_2023}, and remains constant across layers and heads. It is important to note that, 
computing \(Q \times K^T\) requires accessing the K data in KV cache, while computing \(S \times V\) requires only V data. Thus, in the proposed mapping scheme, K(V) refers to either K or V, with identical data layouts but stored in separate blocks.

The proposed KV cache mapping scheme is compatible with both KVNAND-C and KVNAND-D. In KVNAND-C, adopting the superscalar-like HP dataflow is a prerequisite because of the local KV buffer, whereas in KVNAND-D, it can be applied by the large KV buffer on the SoC. 

\textbf{Access-Aware Block Allocation.} 
KV cache generated earlier in decoding is read at every subsequent step, leading to read disturbance comparable to weights and proportional to output length. To mitigate this stress, at the plane level, blocks are randomized across inference requests to balance wear, with counters tracking access and P/E cycles. When limits are reached, data is migrated or refreshed with the flash translation layer (FTL) table updated accordingly. The sequential page order within blocks is preserved for high read speed \cite{lee_aif_2025}.

%% file: contents/5-evaluation.tex
\section{Evaluation}
\label{sc:eval}

\subsection{Methodology}

\textbf{Baseline and Configuration.} 
We evaluate the following systems: (1) Base-1 (Weight-only IFC + DRAM): IFC design storing all model weights in NAND flash, while KV cache resides entirely in DRAM. This architecture is consistent with existing IFC-based on-device LLM accelerators. We scale the Lincoln architecture \cite{sun_lincoln_2025}, configured with one IFC die and one DRAM attached to each of the 8 LPDDR channels to ensure that the DRAM has sufficient capacity to store long text sequences. (2) Base-2 (Naive KV-in-Flash): Derived from Base-1 by simply replacing DRAM with common NAND flash for KV cache storage. (3) KVNAND-D-(G1+G2): Proposed discrete variant storing model weights in one group of IFC dies (G1) and KV cache in another (G2). The suffix (G1+G2) specifies the channel and die allocation. For example, KVNAND-D-(6+2): 8 channels \(\times\) 1 die per channel, with 3 channels assigned to G1 and 1 channel to G2. (4) KVNAND-C-\textit{x}: Proposed compact variant that co-locates weights and KV cache within \textit{x} dies. For a fair comparison, we use the KVNAND-C-16 configuration.

KVNAND-D and KVNAND-C extend Base-2 by replacing the KV cache flash with compute-enabled IFC dies. This maintains the same number of memory dies in each channel, an advantage of eliminating DRAM, since each channel can instead host additional IFC dies. As shown later in the scalability analysis, even with the same number of IFC dies as Base-1, completely removing DRAM and adding flash capacity can still deliver comparable performance.

\begin{scriptsize}
\begin{table}[!t]
\renewcommand{\arraystretch}{1.3}
\vspace{5pt}


\centering
\caption{Configuration of KVNAND and baseline}
\label{table:table1}
{\fontsize{8pt}{8.5pt}\selectfont 

\begin{minipage}{1\linewidth}
\centering
\begin{tabularx}{\linewidth}{%
  >{\centering\arraybackslash}m{1.2cm}|%
  >{\centering\arraybackslash}X
}
\multicolumn{2}{c}{\textbf{NPU Configuration}} \\ \hline
Compute & 32 TOPS (BF16 Tensor Core), 4.60 W \\ \hline
SRAM    & 5MB KV buffers for KVNAND-D, 0.36 W \\ \hline
\end{tabularx}
\end{minipage}


\begin{minipage}{\linewidth}
\centering
\renewcommand{\arraystretch}{1.2}
\begin{tabularx}{\linewidth}{>{\centering\arraybackslash}X}
\multicolumn{1}{>{\centering\arraybackslash}X}{\textbf{Memory System Configuration}} \\ \hline
DRAM: LPDDR5X, 16Gb, 8GB/s, 7\,pJ/bit 
(Base-1)  \\
NAND Flash: ONFI 6.0, 128Gb, 4.8GB/s, 4.9\,pJ/bit\\
\hline
\end{tabularx}
\end{minipage}


\begin{minipage}{\linewidth}
\centering
\renewcommand{\arraystretch}{1.2}
\begin{tabularx}{\linewidth}{>{\centering\arraybackslash}m{1.2cm}|>{\centering\arraybackslash}X}
\multicolumn{2}{c}{\textbf{KVNAND Flash Configuration}} \\ \hline
\multirow{4}{=}{\makecell[c]{\textbf{Flash}\\\textbf{layer}}}
& SLC, 96 wordline-layers, (4KB + 448B ECC) / page \\
& 768 pages / block, 177 blocks / plane, 32 planes \\
& 4.1484\,Gb per plane, 132.75\,Gb per die \\
\cline{2-2}  
& tR = 4\,$\mu$s, tP = 75\,$\mu$s, 3\,pJ/bit (read), 7.5\,pJ/bit (program)\\ \cline{2-2}  
\hline
\multirow{6}{=}{\makecell[c]{\textbf{Logic}\\\textbf{layer}}}
& 16 FMACs + 8\,KB KV buffer per plane for KVNAND-C \\
& 0.0737\,mm$^2$, 6.98\,mW per plane \\ \cline{2-2}
& ECC: BCH(9088, 8192, 64), 14.6\,Gb/s per plane \\ 
& dec/enc: 0.1478/0.0598\,mm$^2$, 5.24/1.2\,mW  per plane\\ \cline{2-2}
& 260\,KB Global buffer \\
& 0.4095\,mm$^2$, 18.4\,mW in total \\ \cline{2-2}
\hline
\end{tabularx}
\end{minipage}
\vspace{-10pt}

} 

\end{table}
\end{scriptsize}

Table \ref{table:table1} lists hardware parameters used in our evaluation
Most parameters are derived from Lincoln \cite{sun_lincoln_2025} for consistency and fair comparison. Performance modeling follows Cambricon-LLM \cite{yu_cambricon-llm_2024} with SSDsim \cite{hu_achieving_2010}  to capture both memory and compute behaviors of IFC subsystem, while NVSim-based \cite{dong_nvsim_2012}  3DFPIM simulators \cite{lee_3d-fpim_2022} validate flash latency/energy using conservative estimates. 
The digital logic of the IFC layer, with reduced FMAC units of Lincoln \cite{sun_lincoln_2025}, is synthesized using Design Compiler and also scaled to 20 nm, while on-chip SRAM buffers are generated using the ARM memory compiler.
Considering area limits, KVNAND-C provisions 8 KB KV buffers per plane, while KVNAND-D employs a 5 MB SoC-side KV buffer. DRAM access energy in Base-1 is modeled at 7.0 pJ/bit \cite{dong_nvsim_2012,noauthor_keynote_2019}, flash interface access at 4.9 pJ/bit \cite{na_18-gbspin_2021}, and internal flash reads at 3 pJ/bit \cite{sun_lincoln_2025}. Write operations are modeled with \(tP = 75 \,\mu s\) \cite{kouchi_135_2020} and 7.5 pJ/bit energy (\(\approx 2.5\times\) read energy \cite{kouchi_135_2020}).
We evaluate MoE FFN using TP, which shards the expert weights across the IFC dies and executes the low-overhead router on the NPU.

\textbf{Benchmarks.} 
We evaluate five full-precision LLMs (MHA-based OPT-30B \cite{zhang_opt_2022}, LLaMA2-7B \cite{touvron_llama_2023-1} and GQA-based LLaMA3.1-8B \cite{llamateam_llama_2024}, LLaMA3.1-70B \cite{llamateam_llama_2024}, Mixtral8×7B \cite{jiang_mixtral_2024}) spanning different scales and architectures and various mixed-precision quantization schemes for comprehensive DSE. 

\subsection{End-to-end Performance and Energy Efficiency}


\begin{figure}[tb]
    \centering
    \includegraphics[width=1.0\linewidth, trim=30 30 15 15, clip]{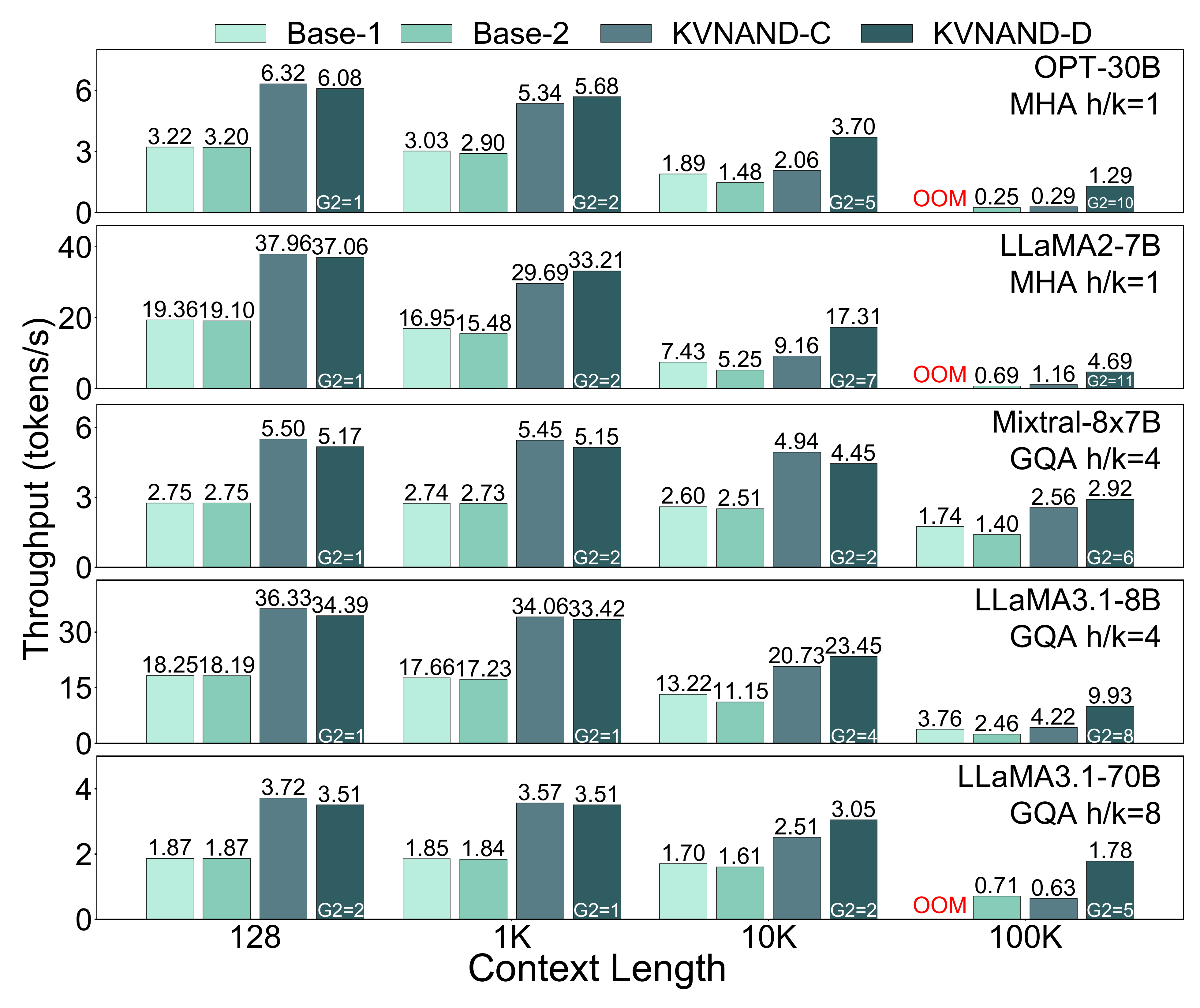}
    \caption{Throughput of Base-1, Base-2, KVNAND-C, and KVNAND-D under 1K, 10K, 100K context length.}
    \label{fig:token_per_sec_comparison_combined}
    \vspace{-10pt}
\end{figure}

\begin{figure}[tb]
    \centering
    \includegraphics[width=1.0\linewidth, trim=4 5 5 5, clip]{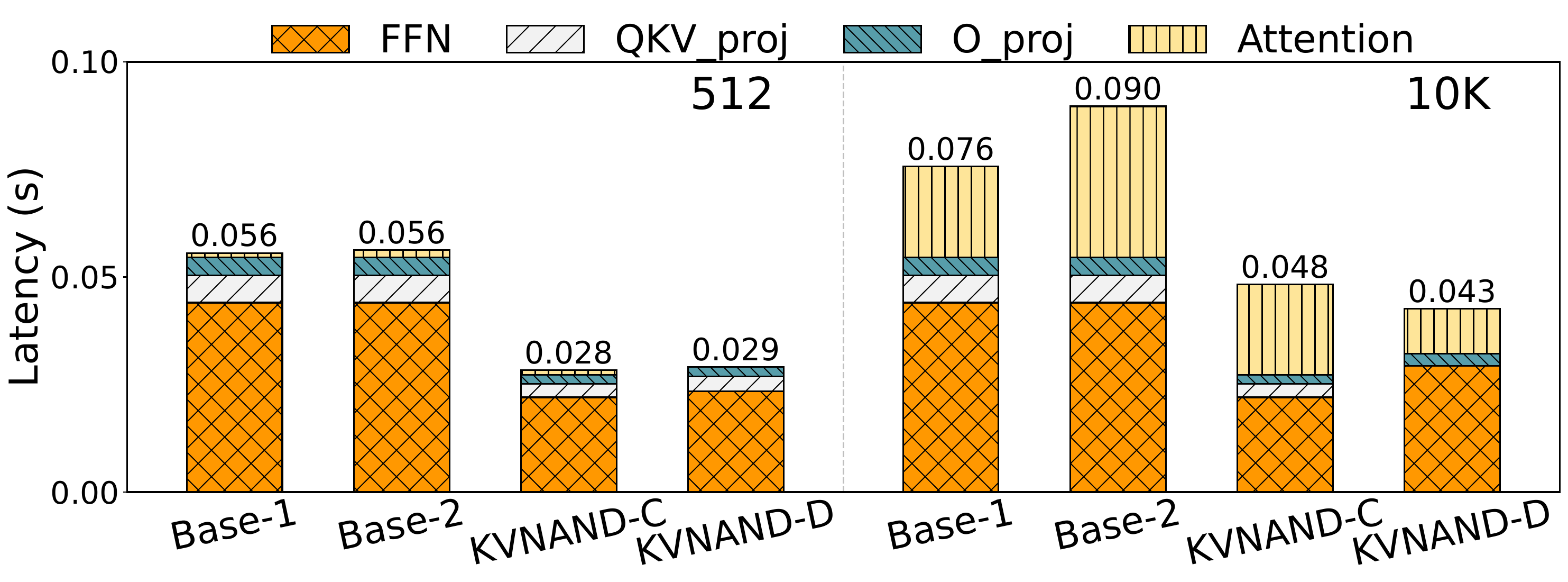}
    \caption{Latency breakdown comparison (LLaMA3.1-8B).}
    \label{fig:inference_time_breakdown_v4}
    \vspace{-1.6pt}
\end{figure}

\textbf{Performance Speedup.}
Figure \ref{fig:token_per_sec_comparison_combined} compares the decoding throughput of KVNAND-D and KVNAND-C against two baselines, while Figure \ref{fig:inference_time_breakdown_v4} provides a breakdown of the major components of decode latency for the 8B model under 1K and 10K context lengths. To ensure optimal benefits, the hardware configuration for KVNAND-D is selected based on the results of our DSE, which is presented subsequently. Three key observations can be made:


First, KVNAND-D consistently improves throughput across most models and context lengths. At a 100K-token context, it achieves 5.2\(\times\), 6.8\(\times\), 4.0\(\times\), 2.5\(\times\), 2.1\(\times\) higher decode speed than Base-2 for the OPT-30B, LLaMA2-7B, LLaMA3.1-8B, LLaMA3.1-70B, and Mixtral-8×7B, respectively. The performance gains become more pronounced as context length increases, driven by the larger ratio of KV cache–related attention computations as depicted in Figure 13. Although GQA's higher arithmetic intensity limits the IFC speedup, our design remains beneficial thanks to the additional computational power that is incorporated based on our roofline intersection analysis.

Second, KVNAND-C demonstrates higher gains at shorter context lengths. At 128 context length, it achieves 1.98\(\times\) and 1.05\(\times\) geomean speedup over Base-1 and KVNAND-D across all models. This advantage stems from its higher degree of parallelism in accelerating FFN computations as depicted in Figure \ref{fig:inference_time_breakdown_v4}. At shorter context length, the workload is FFN-bound, and the optimal DSE split skews heavily towards G1 (e.g., G1=15, G2=1 for LLaMA3.1-8B, 1K context-length). This large G1 allocation makes its FFN computational power comparable to that of KVNAND-C, resulting in similar FFN performance. However, for longer contexts, more frequent KV cache transfers amplify the limitation of its smaller KV buffer, diminishing performance benefits relative to KVNAND-D. 



Finally, we observe that Base-1 experiences an out-of-memory (OOM) failure at the 100K context length, as its DRAM capacity is insufficient to accommodate the requisite KV cache. Even the three GQA models, which are more KV-efficient, already exhaust the DRAM capacity at context lengths of $~$50K. Base-2 (naive KV-in-Flash) suffers severe slowdowns due to flash access, resulting in significant additional latency from KV cache transfers during attention computation. This problem becomes more pronounced in longer contexts and highlights the necessity of KVNAND.

\begin{figure}[tb]
    \centering
    \includegraphics[width=1.0\linewidth, trim=4 5 5 5, clip]{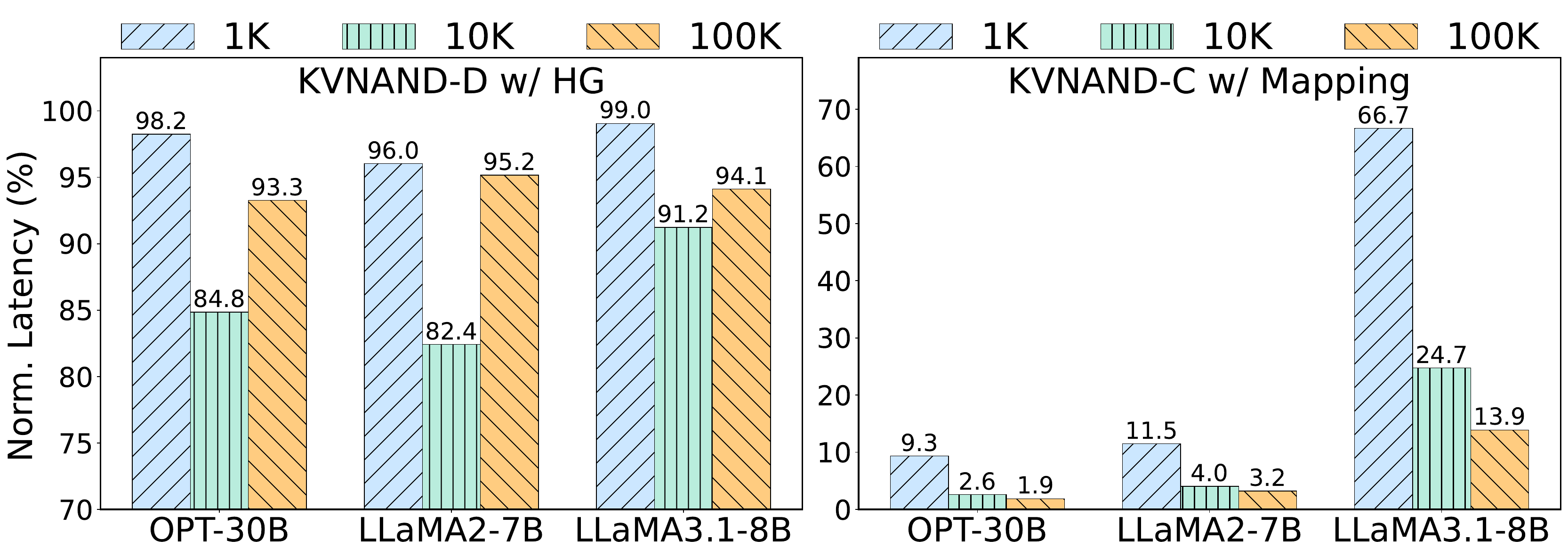}
    \caption{Ablation study of HG parallelism in KVNAND-D and KV cache mapping scheme in KVNAND-C.}
    \label{fig:ablation_time_comparison_left_right}
    \vspace{-10pt}
\end{figure}

\textbf{Ablation Study.} 
Figure \ref{fig:ablation_time_comparison_left_right} evaluates the impact of two key optimizations. For KVNAND-D, HG parallelism reduces latency across different context lengths, normalized to a baseline without the dataflow optimization. At 10K tokens, latency drops to as low as 82.4\%. The benefit is maximized when attention latency becomes comparable to QKV generation, enabling effective overlap between the two. However, at short contexts (e.g., 1K), QKV generation dominates latency, while at very long contexts (e.g., 100K), attention computation dominates, making the benefit less pronounced.

For KVNAND-C, the proposed page-level KV cache mapping optimization substantially reduces the portion of attention latency spent on flash reads. The improvement scales with context length, since larger KV caches exacerbate read overhead. At 100K tokens, the latency drops to only 1.9\% for MHA 30B model compared to a design without mapping. The latency gains for GQA models are comparatively diminished. Notably, this optimization yields even greater benefits when applied to KVNAND-D, which provisions a larger KV buffer. In this analysis, we assume that the KV cache corresponding to the evaluated context length is fully stored in flash.


\begin{figure}[tb]
    \centering
    \includegraphics[width=1.0\linewidth, trim=4 5 5 5, clip]{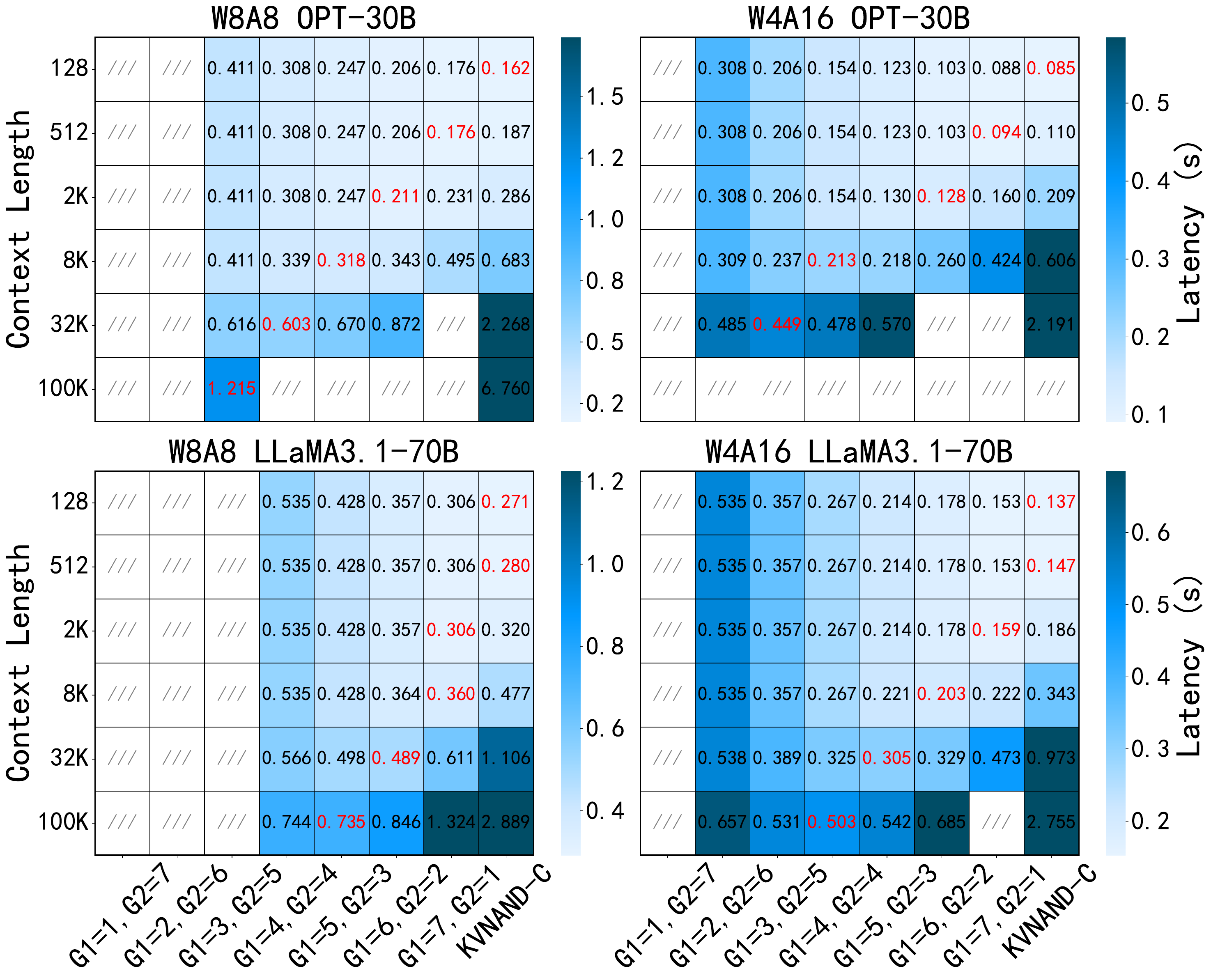}
    \caption{Latency heatmaps for different KVNAND configurations across model scales, quantization schemes, and context lengths. A blank space indicates OOM.}
    \label{fig:combined_models_heatmaps}
    \vspace{-10pt}
\end{figure}

\textbf{Scalability Study and Design Space Exploration (DSE).}
As shown in Figure \ref{fig:combined_models_heatmaps}, we further scale down KVNAND to an 8-die IFC configuration and explore various grouping schemes for KVNAND-D, where the number of dies in G1 ranges from 1 to 7. The study is conducted on both the 30B MHA and 70B GQA models under two widely adopted quantization settings: W8A8 (8-bit weight and 8-bit activation/KV cache) and W4A16 (4-bit weight and 16-bit activation/KV cache) \cite{yu_cambricon-llm_2024}. Each grid cell in the heatmaps represents the inference latency under different sequence lengths, with red values indicating the best performance. All configurations must also satisfy storage constraints; i.e., the combined size of weights and KV cache must not exceed the available flash capacity. Configurations that violate this constraint are marked as OOM and denoted by blank cells. This OOM constraint is triggered by two primary factors. First, MHA models are highly susceptible to long context lengths due to their large KV cache footprint. Separately, large-scale models with their massive weight parameters are incompatible with configurations that provision too few G1 dies.
Our evaluation targets system-level trade-offs of KVNAND, following prior PIM studies \cite{li_h2_2025}, instead of RTL simulation, to enable efficient DSE in the early design stage. This DSE-driven approach also enables adaptive on-device, software-defined reconfiguration of KVNAND. When the workload evolves (e.g., model updating or context-length mode switching), the system can be dynamically re-configured by the DSE results to apply the optimal settings. Key observations of our DSE are as follows.

First, the benefit of balanced grouping. For both 30B and 70B models, balanced die allocation between G1 and G2, or adopting the KVNAND-C scheme, generally achieves the lowest latency across a wide range of sequence lengths. For instance, in LLaMA3.1-70B (W4A16), as the required context length increases and attention computation dominates, KVNAND-D gradually outperforms KVNAND-C beyond 2K tokens. In this regime, allocating more IFC dies to G2 for KV storage and attention computation improves performance, with the optimal configuration reaching 4 dies in G2 at 100K tokens. Conversely, in shorter contexts where FFN dominates the latency, KVNAND-C or allocating more dies to G1 is more advantageous. This effect is more pronounced under GQA model, since the larger weight size and smaller KV cache size shift the bottleneck toward weight-related GEMVs. 

Second, the impact of quantization. At the same model size and context length, W8A8 quantization favors allocating more dies to G1 due to the larger weight size, while W4A16 favors allocating more dies to G2 as activation and KV cache handling become the dominant bottleneck. In KVNAND designs, W4A16 achieves better performance than W8A8, in contrast to previous work where attention is calculated on the NPU \cite{yu_cambricon-llm_2024} due to IFC-accelerated attention.

Third, scaling with sequence length. At short contexts (\(\le5K\)), performance is relatively insensitive to configuration because KV cache overhead is small and weight-related GEMVs dominate. As sequence length increases to 50K–100K, latency rises sharply for unbalanced groupings due to intensified KV cache traffic. This highlights the importance of tuning the G1/G2 allocation according to target context length.

\begin{figure}[tb]
    \centering
    \includegraphics[width=1.0\linewidth, trim=4 5 5 5, clip]{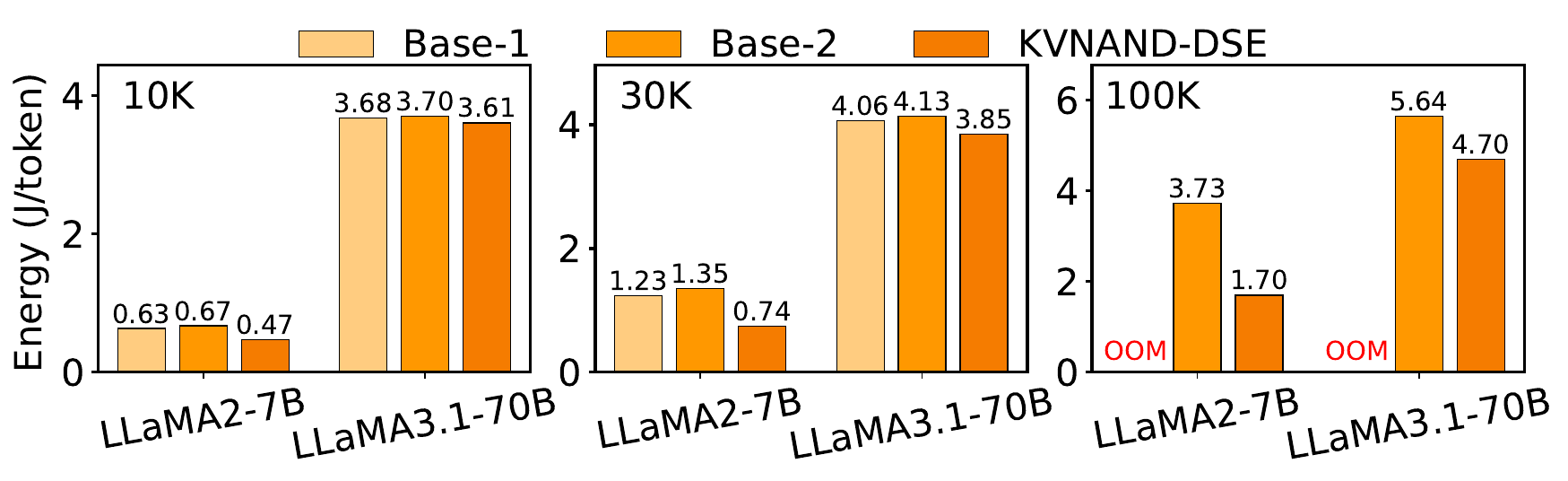}
    \caption{Energy consumption of per-token decoding. }
    \label{fig:energy_comparison_combined}
    \vspace{-10pt}
\end{figure}

\vspace{-4pt}
\begin{tcolorbox}[colback=white,       
boxsep=0.5mm,
  colframe=black,        
  boxrule=0.8pt,         
  arc=0.5mm,   
  left=1mm,              
  right=1mm,             
  top=1mm,               
  bottom=1mm             
  ]
\textbf{\emph{Takeaway 1:}} Balanced die allocation is key. KVNAND-D is better suited for long-context applications, offering higher performance and reliability after careful design-space configuration, while KVNAND-C is more favorable at shorter contexts.

\textbf{\emph{Takeaway 2:}} Quantization shifts bottlenecks. W4A16 favors G2-heavy allocations and yields better performance under IFC-accelerated attention, whereas W8A8 favors G1-heavy allocations due to larger weight size.

\end{tcolorbox}
\vspace{-4pt}



\textbf{Energy Efficiency.}
Figure 16 reports the energy consumption for two representative models under different context lengths. Overall, KVNAND after the proposed DSE achieves consistently lower energy consumption than both baselines, with the advantage becoming more pronounced at longer contexts. At 10K tokens, KVNAND reduces energy consumption to 0.75\(\times\) and 0.98\(\times\) of Base-1 for the LLaMA2-7B and LLaMA3.1-70B models, respectively, and at 100K tokens, the savings further improve to 0.46\(\times\) and 0.83\(\times\) of Base-2. These reductions stem from significantly less data movement: IFC computation eliminates repeated transfers of KV cache data. GQA models inherently reduce energy consumption for long-context inference owing to smaller KV head numbers.

KVNAND-D still requires some external data transfers, but the overall volume is much smaller due to offloaded QKV generation and reduced intermediate data movement. In addition, flash write energy is amortized across multiple accumulated updates, keeping the overall cost modest.

In terms of thermal throttling, prior analysis of 3D-stacked IFC dies \cite{sun_lincoln_2025} has shown that their power density is acceptable. KVNAND adopts the same flash configuration and maintains comparable compute resources on the logic die, so its thermal profile is expected to remain similar. The additional buffer required in KVNAND-C introduces only a minor overhead as listed in Table 1, confirming that thermal impact is negligible.


\subsection{Cost Analysis}
The primary cost components of KVNAND come from SLC-mode 3D NAND flash, hybrid bonding, and peripheral dies. We base our estimation on YTMC’s 128-layer TLC flash (8.5 Gb/mm²) \cite{techpowerup_zhitai_2025}, which already includes hybrid bonding and peripheral die costs, priced at \$0.11/GB. Converting to SLC mode (1.8 Gb/mm² density) results in \$0.52/GB. Considering the 1.22\(\times\) area and 2\(\times\) hybrid bonding I/O per die required for page buffers, the yield can be conservatively estimated as 58\% (normal, 80\% yield base), leading to \$0.72/GB \cite{sun_lincoln_2025}.

For 8-channel, 8-die configuration of KVNAND-D-(4+4), flash cost is approximately \$92.16 for combined weight and KV cache storage. This is significantly lower than DRAM-based IFC designs; for the same model and context-length KV cache capacity, replacing flash with LPDDR5 (\(\$4.62\)/GB \cite{mouser_electronics_sdram_nodate}) would increase the memory cost by more than \(2\times\) (\$295.68).

The additional logic in IFC-enabled KV dies incurs minimal area overhead (\( < 3\% \) per die) and negligible cost compared to the flash array. While KVNAND requires both weight and KV dies to integrate IFC logic, this reuse of the existing peripheral die layout ensures that the incremental manufacturing cost remains small. The adoption of multi-die packages and HB interconnects follows established 3D NAND manufacturing practices, and the associated packaging cost is estimated to be \(<15\%\) of raw die cost \cite{feng_chiplet_2022}.

\subsection{Reliability Evaluation}
To evaluate the long-term reliability, we consider a 65B-parameter model \cite{touvron_llama_2023-1} running on KVNAND at 3 tokens/s throughput. Under the pessimistic assumption of continuous decoding for 5 years, the total KV cache generated amounts to approximately 143 TB. For the 8-die, 16 GB configuration of KVNAND-D, which supports up to \(\sim \)50K context length, this corresponds to about 1K P/E cycles, well below the nominal endurance budget of 100K P/E cycles in SLC flash \cite{noauthor_new_2022}. Conservatively, we leave roughly 50K cycles as usable margin for page-read disturbance refresh.

We model the flash-based KV cache read/write process and track the cumulative page-read counts across blocks. At a 50K context length (with 1K input and 1K output tokens per request), the total page read counts per block can reach \(4\times10^6\), which exceeds the intrinsic disturbance limit of \(10^6\) \cite{zambelli_uniform_2017}. However, KVNAND's data mapping and parallelization schemes substantially mitigate this stress. In KVNAND-C-16, head-parallel KV generation reduces maximum per-block page-read counts by approximately \(128\times\) (\(\sim \frac{ k\times page\_size}{KVbuf\_size}\)). In KVNAND-D, separating weight and KV dies further lowers PGRD amplification, achieving an estimated \(2560\times\) reduction. These reductions provide ample reliability margin for sustained long-context inference workloads, ensuring that KV cache storage remains robust throughout device lifetime.









%% file: contents/6-related_work.tex
\section{Related Work}
\label{sc:relatedwork}

\textbf{In-Flash Computing for LLMs.} To enable the deployment of large-scale LLMs such as LLaMA-70B on memory-constrained consumer devices, recent research has explored integrating compute capabilities directly into flash storage, a trend commonly referred to as in-flash computing \cite{yu_cambricon-llm_2024,sun_lincoln_2025,lee_aif_2025}. IFC addresses two major bottlenecks of SSD-based offloading \cite{alizadeh_llm_2024, sheng_flexgen_2023,rasley_deepspeed_2020}: the limited transmission bandwidth between flash and accelerators, and the low internal read throughput of flash arrays. The common principle is to offload memory-bound GEMV operations to logic placed close to the flash array. Among them, Lincoln \cite{sun_lincoln_2025} enhances performance with speculative decoding, reducing iteration latency by generating multiple tokens per step.
They further propose a round-robin row distribution layout for weight data, tailored to the distinct compute and memory access characteristics of the NPU and flash. 
Cambricon-LLM \cite{yu_cambricon-llm_2024} strategically partitions GEMV operations, keeping some linear layers on the NPU while mapping others to flash, thereby balancing channel utilization. AiF \cite{lee_aif_2025} further improves throughput at the circuit level by exploiting the sequential access pattern of weights, introducing charge-recycling reads that bypass conventional wordline precharge/discharge cycles to reduce read latency. These techniques are orthogonal to our work and can be directly incorporated into KVNAND. Distinct from prior IFC architectures, KVNAND offloads the KV cache to flash, eliminating the high cost and energy overhead of DRAM for KV storage and transfers. Even on devices that inevitably integrate DRAM for system software and embeddings, KVNAND remains applicable and reduces DRAM pressure, enabling efficient long-context LLM inference.
In particular, KVNAND-D can leverage available DRAM as a large KV buffer, reducing on-die SRAM requirements.

\textbf{LLM Acceleration.} Numerous quantization-based techniques \cite{dettmers2023qlora, frantar2022gptq, guo2023olive,lin2024awq,liu2023llm,yuan2023rptq,xiao2023smoothquant,guan2024aptq,wang2023bitnet} have been proposed to reduce model size and enable LLM deployment on edge devices. These methods are orthogonal to KVNAND and are incorporated into our DSE framework as input configurations for design optimization. Beyond quantization, sparse attention has emerged as another widely adopted approach to lower KV cache access and computation demand \cite{chaudhari2021attentive,chen2021scatterbrain,liu2023scissorhands,liu2023deja,ribar2023sparq,xiao2023efficient,zhang2023h2o,tang2024razorattention,sun2024shadowkv,li2024survey}. Sparse attention is based on the observation that not all tokens contribute equally to the final attention output.
Determining which tokens to prune often introduces algorithmic complexity and potential accuracy degradation. 
Building on its ability to mitigate KV cache fragmentation, the KVNAND architecture has the potential to support diverse sparsity schemes as well.

%% file: contents/7-conclusion.tex
\section{Conclusion}
\label{sc:con}
This paper presents KVNAND, a DRAM-free NPU–IFC architecture that stores both model weights and the KV cache in compute-enabled NAND flash. By co-designing dataflow, head-group parallelism, and page-level KV cache mapping, KVNAND effectively addresses the challenges of KV storage in flash. Evaluation across diverse models and context lengths shows that KVNAND achieves up to 2.3\(\times\) speedup and 0.75\(\times\) energy consumption relative to SoTA DRAM-based IFC designs at 10K context length compared to baseline. Furthermore, KVNAND fully resolves the OOM barrier at 100K context lengths, delivering an inference throughput of 10 tokens/s on LLaMA3.1-8B. Beyond performance, the low cost and energy efficiency make KVNAND a critical enabling technology for deploying long-context LLMs on edge devices. We hope KVNAND motivates further exploration of IFC for next-generation LLM systems.